\documentclass{article}
\usepackage[utf8]{inputenc}
\usepackage{float}
\usepackage{jheppub}
\usepackage{amsmath}
\usepackage[english]{babel}
\usepackage{amssymb}
\usepackage{mathtools}
\usepackage{color}
\definecolor{ao}{rgb}{0.0, 0.5, 0.0}

\usepackage{graphicx}
\usepackage{subcaption}
\usepackage{longtable}
\usepackage{tikz}
\usetikzlibrary{positioning,calc}
\usetikzlibrary{backgrounds}
\tikzset{
  node in/.style={
    circle,
    draw=black,
    fill=gray!10,
    minimum size=18pt,
    inner sep=1pt
  },
  node hidden/.style={
    rectangle,
    draw=black,
    fill=red!10,
    minimum size=18pt,
    inner sep=1pt
  },
  connect/.style={
    -, thick
  },
  gfield/.style={
    ->,
    dashed,
    gray,
    thick
  }
}

\usetikzlibrary{arrows.meta}

\usepackage{comment}
\usepackage{capt-of}
\usepackage{amsthm}
\usepackage{hyperref}
\hypersetup{
    colorlinks=true,
    linkcolor=blue
    }

\usepackage{bm}

\usepackage{caption}
\usepackage{subcaption}

\allowdisplaybreaks

\def\0{\mbox{\tiny $0$}}
\def\1{\mbox{\tiny $1$}}
\def\2{\mbox{\tiny $2$}}
\def\3{\mbox{\tiny $3$}}
\def\4{\mbox{\tiny $4$}}
\def\5{\mbox{\tiny $5$}}
\def\6{\mbox{\tiny $6$}}
\def\7{\mbox{\tiny $7$}}
\def\8{\mbox{\tiny $8$}}
\def\9{\mbox{\tiny $9$}}

\def\k{k_{_{B}}}

\def\r{\rangle}
\def\R{\mathcal{R}}
\def\l{\langle}
\def\m{\bar{m}}

\def\q{\bar{q}}
\def\n{\bar{n}}

\def\b{\beta^{'}}

\newcommand{\SOMMA}[2]{\displaystyle\sum\limits_{#1}^{#2}}

\long\def \beq#1\eeq {\begin{equation} #1 \end{equation}}
\long\def \beaq#1\eeaq {\begin{equation}\begin{aligned} #1 \end{aligned}\end{equation}}
\long\def \bes#1\ees {\begin{equation}\begin{split} #1 \end{split} \end{equation}}
\long\def \bea#1\eea {\begin{eqnarray} #1 \end{eqnarray}}
\long\def \bse[#1]#2\ese {\begin{subequations}\label{#1}\begin{align} #2 \end{align}\end{subequations}}

\title{Dense Associative Memory with biased patterns: a Replica Symmetric analysis}
\author[1,2]{Linda Albanese}
\author[1,2]{, Andrea Alessandrelli}
\author[3,4]{, Federico Carella}

\affiliation[1]{Dipartimento di Matematica e Fisica ``Ennio De Giorgi'', Universit\`a del Salento, Lecce, Italy.}
\affiliation[2]{Istituto Nazionale di Fisica Nucleare, Sezione di Lecce, Italy.}
\affiliation[3]{Now at Fondazione CMCC - Centro Euro-Mediterraneo sui Cambiamenti Climatici, Lecce, Italy.}
\affiliation[4]{Dipartimento di Fisica e Astronomia “Augusto Righi” - DIFA, Universit\`a di Bologna, Bologna, Italy}

\abstract{
We investigate dense higher-order associative memories in the high storage regime when the stored patterns are biased, namely when the entries of the patterns are not symmetrically distributed around zero. In this setting, the standard Hebbian prescription must be modified by recentering and rescaling the pattern entries, and an additional term must be introduced in the Hamiltonian to enforce consistency between the average activity of the network and that of the stored patterns. As a first step, we perform a signal-to-noise analysis in the zero-temperature limit and show that the bias reduces the effective storage capacity through a multiplicative correction factor $(1-b^2)^P$, while preserving the superlinear scaling with the system size. We then derive the quenched statistical pressure within the Replica Symmetric framework by means of Guerra's interpolation method and obtain the corresponding self consistency equations for the relevant order parameters. The analytical treatment confirms the heuristic prediction of the signal-to-noise argument, showing that the same bias dependent renormalization naturally emerges in the variance of the cross-talk noise. Finally, we discuss the resulting phase behavior of the model and its implications for retrieval performance in the model.
}

\begin{document}

\maketitle

\section{Introduction}

Associative memory models constitute one of the most paradigmatic frameworks for understanding collective information processing in neural networks and disordered systems. Since Hopfield’s seminal contribution \cite{Hopfield}, these models have provided a simple yet powerful account of how a network of interacting binary units can store a collection of patterns as stable attractors of the dynamics and retrieve them from partial or corrupted inputs. Beyond their original neuroscientific motivation, associative memories have come to play a central role in statistical mechanics \cite{Hopfield, AGS}, machine learning \cite{ramsauer2020hopfield}, and the study of complex systems \cite{peretto1984collective, agliari2019dreaming, fachechi2019dreaming}, where they serve as a natural bridge between spin glass methodologies and information processing \cite{MPV, nishimori2001statistical}.

A significant line of research in this area focuses on enhancing storage and retrieval performance by introducing higher-order interactions. In dense $P-$spin generalisations of the Hopfield model \cite{HopKro1, Krotov2018, Albanese2021}, standard pairwise Hebbian couplings are replaced by interaction tensors of a higher order, enabling the network to encode correlations between multiple points among the stored patterns. Such architectures are known to display a relevant increase in storage capacity \cite{Baldi}. Whereas the classical Hopfield model can only store a number of patterns that scales linearly with system size $N$, dense higher-order models can operate in a high-storage regime, where the number of stored patterns $K$ increases as $K \sim O( N^{P-1})$ \cite{Bovier, Baldi}. This makes them a natural framework for investigating associative memory beyond the conventional low storage regime \cite{Gardner, unsup, super, albanese2024replica}.

However, most of the existing Literature assumes that the stored patterns are unbiased, meaning their entries are independent, identically distributed and symmetric Rademacher variables \cite{Hopfield, AGS}. Under this assumption, the standard Hebbian learning rule is naturally centred and the retrieval problem can be analysed using a well established theoretical framework \cite{Gardner}.
Anyhow, in many realistic scenarios, the statistics of the stored information are not balanced \cite{agliari2022storing, amit1987information, lowe1999storage}. Patterns may exhibit non-zero mean activity, meaning the probabilities of entries taking $+1$ or $-1$ values are no longer equal. This creates an intrinsic bias in the stored memories, influencing both the learning rule and the retrieval dynamics. The presence of biased patterns gives rise to several non trivial issues. Firstly, the Hebbian interaction tensor must be appropriately centred and scaled to eliminate the contribution arising from the non-zero mean and retain the relevant fluctuations \cite{tsodyks1988enhanced}. Secondly, even after this transformation, the network dynamics may not converge to configurations that exhibit the same average activity as the stored patterns. The issue can be addressed by adding an extra constraint term to the Hamiltonian \cite{amit1987information, agliari2022storing}. This soft mechanism makes sure the desired level of average activation is reached. Therefore, the resulting model depends not only on the storage load and temperature, but also on the degree of bias and the strength of the activation constraint.

\par\medskip
In this study, we examine a dense higher-order Hopfield model featuring biased patterns within the high storage regime. We analyse the thermodynamic behaviour of the system within the Replica Symmetric ansatz and our primary objective is to understand the influence of pattern bias on both the retrieval threshold and the storage capacity. To this end, we first conduct a signal-to-noise analysis at zero temperature. This heuristic approach suggests that, while there is a bias dependent reduction in the effective load, the dense architecture preserves its superlinear storage capacity, captured by a multiplicative factor $(1-b^2)^P$, with $b$ patterns bias. In this way we show that bias induces a universal renormalization of the effective load, preserving the superlinear scaling while quantitatively modifying the retrieval threshold. We then perform a full thermodynamic analysis based on Guerra’s interpolation method \cite{guerra_broken}. This approach allows us to derive the quenched statistical pressure of the model in the high storage regime, as well as obtain the corresponding self consistency equations for the relevant order parameters: the Mattis magnetization, the mean activity and the two-replicas overlap. The analytical solution corroborates the heuristic prediction obtained from the signal-to-noise analysis: the same factor $(1-b^2)^P$ re-emerges as a renormalisation of the crosstalk noise generated by the non-condensed patterns. This agreement provides a consistent picture of how bias quantitatively affects retrieval performance while maintaining the qualitative scaling behaviour of the model.

\par\medskip
The paper is organized as follows. In Section~\ref{sec:model}, we define the model and introduce the biased pattern distribution together with the corresponding recentered Hebbian prescription and activity constraint. In Section~\ref{sec:signal_to_noise}, we present the signal-to-noise analysis and derive the associated estimate for the retrieval threshold at zero temperature. In Section~\ref{sec:guerra}, we develop the Guerra interpolation scheme and derive the RS expression of the quenched free energy, along with the self consistency equations for the order parameters and we discuss the resulting phase diagram and the retrieval properties of the model as functions of the bias, the storage load, the temperature, and the strength of the activation constraint.

\section{Generalities}
\label{sec:model}

In this section we introduce the model under analysis, namely the \textit{dense biased associative memory}. 

Before doing this, let us analyze the \textit{standard} dense generalization of associative memory with Hebbian synaptic couplings \cite{Gardner, Albanese2021}. In this case, one considers a collection of $K$ Boolean patterns of length $N$, denoted by $\{\boldsymbol{\xi}^{\mu}\}_{\mu=1}^K$, whose entries are i.i.d.\ Rademacher random variables, namely
\begin{equation}
\mathcal{P}(\xi_{i}^{\mu})
=\frac{1}{2}\delta(\xi_{i}^{\mu}-1)+\frac{1}{2}\delta(\xi_{i}^{\mu}+1),
\qquad \forall\,i=1,\dots,N,\quad \mu=1,\dots,K.
\label{eq:patterns_general}
\end{equation}
These random variables are used to construct the $P$-body Hebbian interaction tensor
\begin{equation}
J^{(P)}_{i_1,\dots,i_P}(\boldsymbol\xi)
=\frac{1}{N^{P-1}}\sum_{\mu=1}^{K}\xi_{i_1}^\mu\cdots\xi_{i_P}^\mu,
\label{eq:P_Hebbian_standard}
\end{equation}
which encodes the $P$-points correlations among the stored patterns and models the effective interactions among neurons in the network.

The neural configuration is described by
\begin{equation}
\boldsymbol{\sigma} = (\sigma_1,\dots,\sigma_N)\in \Omega,
\qquad \Omega=\{-1,+1\}^{N},
\end{equation}
and the system is governed by the Hamiltonian
\begin{equation}
H_{N}^{(P)}(\boldsymbol{\sigma}\,|\,\boldsymbol{\xi})
=
-\sum_{(i_1,\dots,i_P)=1}^{N}
J^{(P)}_{i_1\cdots i_P}(\boldsymbol \xi)\,\sigma_{i_1}\cdots\sigma_{i_P},
\label{eq:hamiltonian}
\end{equation}
where the notation $\sum_{(i_1,\dots,i_P)}$ stands for $\sum_{\substack{i_1,\dots,i_P \\ i_1 < \dots < i_P}}$.

The retrieval quality of pattern $\boldsymbol{\xi}^\mu$ is quantified by the corresponding \emph{Mattis magnetization}
\begin{equation}
m_{\mu}(\boldsymbol{\sigma} \vert \boldsymbol \xi)
=\frac{1}{N}\sum_{i=1}^{N}\xi_i^\mu \sigma_i,
\qquad \mu=1,\dots,K.
\label{eq:Mattis}
\end{equation}
It is well known \cite{Gardner,HopKro1} that these higher-order generalizations of the standard Hopfield model can retrieve patterns with almost perfect accuracy even when the number of stored patterns grows superlinearly with the system size, namely
\begin{equation}
K\sim \mathcal{O}(N^{P-1}).
\end{equation}

Despite the extensive Literature on this class of models, the case in which the pattern entries are not balanced has received much less attention. In particular, a systematic investigation is still lacking when the distribution of the pattern entries is biased, i.e., when the probabilities of taking the values $+1$ and $-1$ are no longer equal.

To model this situation, we replace \eqref{eq:patterns_general} with the biased distribution
\begin{equation}
\mathcal{P}(\xi_{i}^{\mu})
=\frac{1+b}{2}\delta(\xi_{i}^{\mu}-1)+\frac{1-b}{2}\delta(\xi_{i}^{\mu}+1),
\qquad \forall\,i=1,\dots,N,\quad \mu=1,\dots,K,
\label{eq:patterns_general_bias}
\end{equation}
where $b\in[-1,+1]$ is a tunable parameter measuring the degree of bias, with the case $b=0$ that encodes the unbiased situation. In the biased case, it is known \cite{Amit} that the Hebbian prescription \eqref{eq:P_Hebbian_standard} must be suitably modified by recentering and rescaling the pattern entries. One therefore introduces
\begin{equation}
J^{(P)}_{i_1,\dots,i_P}
=\frac{1}{N^{P-1}}\sum_{\mu=1}^{K}\eta_{i_1}^\mu\cdots\eta_{i_P}^\mu,
\label{eq:P_Hebbian_bias}
\end{equation}
where
\begin{equation}
\eta_i^\mu = \frac{\xi_i^\mu - b}{\sqrt{1-b^2}}.
\end{equation}
This change is natural since $\mathbb{E}[\eta_i^\mu]=0$ and $\mathbb{E}[(\eta_i^\mu)^2]=1$, so the unbiased variables $\bm\eta$ plays the same role of the unbiased variables in the standard setting, having the same two moments.

\par\medskip
However, this modification alone is not sufficient. Indeed, inserting \eqref{eq:P_Hebbian_bias} into \eqref{eq:hamiltonian} does not guarantee that the neural configuration converges toward one of the original stored patterns, which now have average activation equal to $b$ \cite{amit1987information,agliari2022storing}. To enforce this property, one needs to introduce an additional term in the Hamiltonian, playing the role of a \textit{pseudo-Lagrange multiplier}, in order to constrain the network toward configurations with the same mean activation as the stored patterns. For this reason, the Hamiltonian considered in our analysis is
\begin{align}
H_{g,N}^{(P)}(\boldsymbol{\sigma}\,|\,\boldsymbol{\xi})
=
-\dfrac{1}{N^{P-1}}\sum_{\mu=1}^{K}\sum_{(i_1, \dots, i_P)=1}^{N}
\eta_{i_1}^\mu\cdots\eta_{i_P}^\mu\,\sigma_{i_1}\cdots\sigma_{i_P}
+ \dfrac{g}{2}N\left(\dfrac{1}{N}\sum_{i=1}^{N}\sigma_i - b\right)^2,
\label{eq:hamiltonian_starting}
\end{align}
where $g$ is a tunable parameter controlling the strength of the constraint on the mean neural activity.

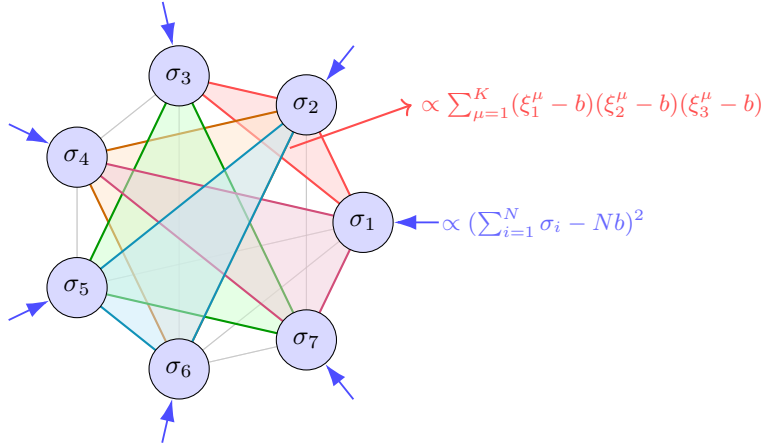
\begin{figure}[t]
\centering
\begin{tikzpicture}[scale=0.8]

\def\n{7}
\def\R{2.5}

\foreach \i in {1,...,7} {
    \coordinate (n\i) at ({360/\n*(\i-1)}:\R);
}

\foreach \i in {1,...,7} {
    \foreach \j in {\i,...,7} {
        \foreach \k in {\j,...,7} {
            \ifnum\i<\j
            \ifnum\j<\k

                \pgfmathtruncatemacro{\keep}{
                    not(
                    (\i==1 && \j==2 && \k==3) ||
                    (\i==2 && \j==4 && \k==6) ||
                    (\i==3 && \j==5 && \k==7) ||
                    (\i==1 && \j==4 && \k==7) ||
                    (\i==2 && \j==5 && \k==6))
                }

                \ifnum\keep=1
                    \draw[gray!60, line width=0.4pt, opacity=0.22]
                        (n\i) -- (n\j) -- (n\k) -- cycle;
                \fi

            \fi
            \fi
        }
    }
}

\filldraw[draw=red!70, fill=red!20, thick, fill opacity=0.5]
    (n1) -- (n2) -- (n3) -- cycle;

\filldraw[draw=orange!80!black, fill=orange!20, thick, fill opacity=0.5]
    (n2) -- (n4) -- (n6) -- cycle;

\filldraw[draw=green!60!black, fill=green!20, thick, fill opacity=0.5]
    (n3) -- (n5) -- (n7) -- cycle;

\filldraw[draw=purple!70, fill=purple!20, thick, fill opacity=0.5]
    (n1) -- (n4) -- (n7) -- cycle;

\filldraw[draw=cyan!70!black, fill=cyan!20, thick, fill opacity=0.5]
    (n2) -- (n5) -- (n6) -- cycle;

\foreach \i in {1,...,7} {
    \node[circle, draw, fill=blue!15, minimum size=8mm] (N\i)
        at (n\i) {$\sigma_{\i}$};
}

\node[align=left, font=\small, red!70, right=1.4cm of n2] (formula)
    {$\propto \sum_{\mu=1}^K (\xi_1^\mu-b)( \xi_2^\mu -b)(\xi_3^\mu-b)$};

\draw[->, red!70, line width=0.8pt]
    ($(n1)!0.33!(n2)!0.33!(n3)$) -- (formula.west);

\node[font=\small, align=left, blue!60] (gterm)
    at ($(n1)+(3.0, 0.0)$)
    {$\propto (\sum_{i=1}^N \sigma_i - Nb)^2$};

\foreach \i [evaluate=\i as \ang using 360/\n*(\i-1)] in {1,...,7} {
    \draw[
        blue!70,
        line width=0.7pt,
        -{Latex[length=3mm,width=2mm]}
    ]
    ($(N\i.\ang)+(\ang:0.75)$) -- (N\i.\ang);
}

\end{tikzpicture}
\caption{Representation of a toy example of the network under investigation with $P=3$ grade of interactions, $N=7$ spins. Vertices correspond to spins $\sigma_i$, and each triangle encodes a term $\sigma_{i_1}\sigma_{i_2}\sigma_{i_3}$ weighted by the biased patterns $\bm \eta^\mu=\bm \xi^\mu - b$. Highlighted triangles represent selected contributions, while gray ones denote the remaining possible interactions. The global term proportional to $ (\sum_{i=1}^N \sigma_i - Nb)^2$ is also indicated. }
\label{fig:toyrepr}
\end{figure}

We have represented a didactic example of how this network works in Fig. \ref{fig:toyrepr}.\\
For later convenience, and in view of the mathematical analysis, we introduce a \textit{modified Mattis magnetization} computed with respect to the new unbiased patterns $\bm \eta$:
\begin{equation}
\tilde{m}_\mu(\boldsymbol\sigma|\boldsymbol\xi)
= \dfrac{1}{N}\sum_{i=1}^{N}\eta_i^\mu \sigma_i = \dfrac{1}{\sqrt{1-b^2}}\Big(m_\mu(\bm\sigma \vert \bm \xi)- b M(\bm\sigma)\Big),
\label{eq:Mattis_eta}
\end{equation}
where we remember that $\eta_i^\mu=\eta_i^\mu(\bm \xi)$ and we have also used the normalized mean activation of the network defined as,
\begin{equation}
M(\boldsymbol\sigma)= \dfrac{1}{N}\sum_{i=1}^{N}\sigma_i.
\label{eq:mean_activation}
\end{equation}
In terms of these order parameters, the Hamiltonian \eqref{eq:hamiltonian_starting} can be rewritten as
\begin{equation}
H_{g,N}^{(P)}(\boldsymbol{\sigma}\,|\,\boldsymbol{\xi})
=
-\frac{N}{2}\sum_{\mu=1}^{K}\big(\tilde{m}_\mu(\boldsymbol{\sigma} \vert \boldsymbol \xi)\big)^P
+\dfrac{g}{2}N \left(M(\boldsymbol\sigma) -b\right)^2.
\label{eq:hamiltonian_magn}
\end{equation}

Writing the Hamiltonian as in \eqref{eq:hamiltonian_magn} makes explicit its linear scaling with the system size $N$ of the network.

As usual, the configuration space $\Omega$ is endowed with the random Boltzmann-Gibbs measure
\begin{equation}
\mathbb{P}^{(P)}_{g,N}(\boldsymbol{\sigma}|\boldsymbol{\xi})
=\frac{\exp\big(-\beta\,H_{g,N}^{(P)}(\boldsymbol{\sigma}\,|\,\boldsymbol{\xi})\big)}
{Z_{g,N}^{(P)}(\beta\,|\,\boldsymbol{\xi})},
\label{BGmeasure}
\end{equation}
where
\begin{equation}
Z_{g,N}^{(P)}(\beta\,|\,\boldsymbol{\xi})
=\sum_{\boldsymbol{\sigma}\in\Omega}
\exp\big[-\beta\,H_{g,N}^{(P)}(\boldsymbol{\sigma}\,|\,\boldsymbol{\xi})\big]
\label{partition-function}
\end{equation}
is the partition function. Accordingly, for any observable $f(\boldsymbol\sigma)$, the Boltzmann-Gibbs average is defined as
\begin{equation}
\omega_{g, N,\boldsymbol\xi}^{(P)}(f(\boldsymbol\sigma))
=
\frac{1}{Z_{g,N}^{(P)}(\beta\,|\,\boldsymbol{\xi})}
\sum_{\boldsymbol{\sigma}\in\Omega} f(\boldsymbol\sigma)\,
\exp\big(-\beta\,H_{g,N}^{(P)}(\boldsymbol{\sigma}\,|\,\boldsymbol{\xi})\big),
\end{equation}
while the quenched average is
\begin{equation}
\label{eq:totalaver}
\langle f(\boldsymbol \sigma) \rangle
=
\mathbb{E}_{\boldsymbol \xi}\Big[\omega_{g, N,\boldsymbol\xi}^{(P)}(f(\boldsymbol\sigma))\Big].
\end{equation}

\par\medskip
Throughout this work, we focus on the \textit{high-storage} regime, namely we assume that the load $\gamma$
\begin{equation}
\gamma = \lim_{N\to\infty}\frac{K}{N^{P-1}} > 0
\label{eq:load}
\end{equation}
remains finite in the thermodynamic limit. Our goal is to study the asymptotic behavior of the corresponding quenched statistical pressure, for fixed $P>2$, under the Replica Symmetric (RS) \textit{ansatz}, namely
\begin{equation}
\mathcal{A}^{(P)}(\beta,g)
=\lim_{N\to\infty}
\frac{1}{N}\mathbb{E}_{\boldsymbol{\xi}}\log Z_{g,N}^{(P)}(\beta\,|\,\boldsymbol{\xi}).
\label{eq:stat-pres-Definition}
\end{equation}
More precisely, we aim to assess the collective retrieval capabilities of the network in the presence of biased patterns and superlinear storage \eqref{eq:load}.

\par\medskip
To characterize the macroscopic behavior of the model, we introduce some relevant functions of the neural configuration $\bm \sigma$, denoted \textit{order parameters}. Retrieval is monitored through the Mattis magnetization with respect to the selected pattern, namely the standard Mattis overlap \eqref{eq:Mattis} and its recentered counterpart \eqref{eq:Mattis_eta}. In addition, we consider the \textit{two-replicas overlap}
\begin{equation}
q_{ab}(\bm \sigma) = \frac{1}{N} \sum_{i=1}^N \sigma_i^a \sigma_i^b,
\label{eq:order_q}
\end{equation}
which quantifies the amount of cross-talk noise affecting the network.
\\
Obtaining an explicit expression for the quenched free energy in terms of the relevant order parameters and control parameters of the theory --namely the bias level $b$, the constraint strength $g$, the inverse temperature $\beta$, and the load $\gamma$-- allows us to derive the associated self consistency equations. Their numerical solution then provides the phase diagrams of the model, i.e., the regions in control parameter space corresponding to different computational behaviors, characterized by the values of the order parameters themselves.

\par\medskip
Before turning to the full thermodynamic analysis, let us present a preliminary and well-known argument based on the signal-to-noise ratio. This approach allows us to assess, with minimal technical machinery and relying essentially on the update rule together with the Central Limit Theorem (CLT), whether in the zero-temperature limit ($\beta\to\infty$) a configuration initially prepared exactly on one of the stored patterns, say $\boldsymbol{\xi}^1$ without loss of generality, remains stable after one step of the dynamics. In other words, we want to determine whether the configuration stays aligned with the retrieved pattern or whether the cumulative interference generated by the other $K-1$ stored patterns is strong enough to drive the system away from it. This argument provides a first estimate of whether the introduction of bias in the pattern distribution modifies the storage capacity of the network.

\section{Signal-to-noise analysis}
\label{sec:signal_to_noise}
Exploiting the mean field structure of the model, the Hamiltonian can be conveniently rewritten as
\begin{equation}
H_{g,N}(\boldsymbol{\sigma}|\boldsymbol{\xi})
=
- \sum_{i=1}^{N} \hat{h}_i(\boldsymbol{\sigma})\,\sigma_i,
\end{equation}
where the effective local field acting on neuron $i$ is given by
\begin{equation}
\label{eq:net_field}
\hat{h}_i(\boldsymbol{\sigma})
:=
\dfrac{P}{2}\dfrac{1}{N^{P-1}}\sum_{\mu=1}^{K}\sum_{\underset{\neq i}{i_2, \dots, i_P=1}}^{N}
\eta_{i}^\mu\eta_{i_2}^\mu\cdots\eta_{i_P}^\mu\,\sigma_{i_2}\cdots\sigma_{i_P}
- g \left(\dfrac{1}{N}\sum_{\underset{\neq i}{j=1}}^{N}\sigma_j - b\right).
\end{equation}

We now introduce a dynamics that drives the system toward lower-energy configurations. In the presence of thermal noise, controlled by the inverse temperature $\beta=T^{-1}\in\mathbb{R}^+$, the neural configuration $\boldsymbol{\sigma}(t)$ at discrete time $t$ evolves synchronously according to
\begin{equation}
\label{eq:update_MC_iclr}
\boldsymbol{\sigma}(t+1)
=
\operatorname{sign}\Big[\tanh\big(\beta\hat{\boldsymbol h}(\boldsymbol{\sigma}(t))\big) + \boldsymbol u(t)\Big],
\end{equation}
where $\hat{\boldsymbol h}$ is the effective field defined in \eqref{eq:net_field}, and
$\boldsymbol u(t)\overset{i.i.d.}{\sim} \mathcal{U}([-1,1]^N)$.


To perform the signal-to-noise analysis, we initialize the network in one of the stored patterns,
\begin{equation}
\label{eq:init}
\boldsymbol \sigma(0)= \boldsymbol\xi^1,
\end{equation}
and compute the effective local field at time $t=0$. We then estimate the overlap between the updated configuration and the same pattern in order to determine whether the initial condition is stable under the dynamics \cite{agliari2020tolerance}.

Since we are interested in the zero-temperature limit, the stochastic update rule \eqref{eq:update_MC_iclr} reduces to the deterministic dynamics
\begin{equation}
\label{eq:update_MC_iclr_ZERO}
\boldsymbol\sigma(t+1)
=
\operatorname{sign}\Big[\hat{\boldsymbol h}(\boldsymbol\sigma(t)) \Big].
\end{equation}
The alignment of the network state with the first pattern can then be monitored through the Mattis magnetization \eqref{eq:Mattis}, which at time $t+1$ reads
\begin{equation}
m_1(t+1)
=
\dfrac{1}{N}\sum_{i=1}^{N}\xi_i^1\operatorname{sign}\Big[\hat{h}_i(\boldsymbol\sigma(t)) \Big]
=
\dfrac{1}{N}\sum_{i=1}^{N}\xi_i^1\sigma_i(t)\operatorname{sign}\Big[\hat{h}_i(\boldsymbol\sigma(t)) \sigma_i(t)\Big].
\end{equation}
Under the initialization \eqref{eq:init}, one has $m_1(0)=1$. Therefore, in order to assess whether the configuration remains perfectly aligned with $\boldsymbol\xi^1$ after one update, we have to evaluate
\begin{equation}
m_1(1)
=
\dfrac{1}{N}\sum_{i=1}^{N}\xi_i^1\sigma_i(0)\operatorname{sign}\Big[\hat{h}_i(\boldsymbol\sigma(0)) \sigma_i(0)\Big]
=
\dfrac{1}{N}\sum_{i=1}^{N}\operatorname{sign}\Big[\hat{h}_i(\boldsymbol\xi^1)\,\xi_i^1\Big].
\end{equation}
Hence, the condition $m_1(1)\approx 1$ can be translated into the requirement that, with high probability,
\begin{equation}
\operatorname{sign}\Big(\hat{h}_i(\boldsymbol\xi^1)\,\xi_i^1\Big)=1,
\end{equation}
or equivalently,
\begin{equation}
\hat{h}_i(\boldsymbol\xi^1)\,\xi_i^1 > 0.
\end{equation}

At this point, the standard signal-to-noise argument consists in decomposing the quantity $\hat{h}_i(\boldsymbol\xi^1)\xi_i^1$ into a deterministic contribution $\mathcal{S}$, associated with the retrieved pattern $\bm \xi^1$, and a fluctuating contribution $\mathcal{N}$, generated by all the other stored patterns:
\begin{equation}
\hat{h}_i(\boldsymbol\xi^1)\xi_i^1=\mathcal{S}+\mathcal{N},
\end{equation}
where
\begin{equation}
\begin{array}{lll}
\mathcal{S}&=&\dfrac{P}{2}\dfrac{1}{N^{P-1}(1-b^2)^{P/2}}\SOMMA{\underset{\neq i}{i_2, \dots, i_P=1}}{N}
(1-b\xi_{i}^1)(1-b\xi_{i_2}^1)\cdots(1-b\xi_{i_P}^1)
- g \left(\dfrac{1}{N}\SOMMA{\underset{\neq i}{j=1}}{N}\xi^1_j - b\right)\xi_i^1,
\\\\
\mathcal{N}&=&\dfrac{P}{2}\dfrac{1}{N^{P-1}(1-b^2)^{P/2}}\SOMMA{\mu>1}{K}\SOMMA{\underset{\neq i}{i_2, \dots, i_P=1}}{N}
(\xi_{i}^1\xi_{i}^\mu-b\xi_{i}^\mu)(\xi_{i_2}^1\xi_{i_2}^\mu-b\xi_{i_2}^\mu)\cdots(\xi_{i_P}^1\xi_{i_P}^\mu-b\xi_{i_P}^\mu).
\end{array}
\end{equation}

The pattern $\boldsymbol\xi^1$ is stable whenever the signal dominates the noise, namely when $|\mathcal{S}| \gtrsim |\mathcal{N}|$. Estimating the signal contribution and the typical size of the noise term by standard probabilistic arguments, together with the CLT, one finds that stability requires
\begin{equation}
\dfrac{2}{P}\dfrac{K}{N^{P-1}(1-b^2)^P}\sim\mathcal{O}(1).
\end{equation}
Equivalently, in terms of the storage load and the bias parameter, this condition becomes
\begin{equation}
K \sim \dfrac{P}{2}(1-b^2)^P N^{P-1}.
\label{eq:load_bias_corr}
\end{equation}

This result shows that the dense higher-order architecture still retains a superlinear storage capacity with respect to the system size. The effect of the bias is not to change the scaling with $N$, but rather to introduce a multiplicative correction factor, $(1-b^2)^P$, which reduces the effective capacity as the bias increases.

\section{Guerra's interpolation technique}
\label{sec:guerra}

The signal-to-noise argument presented in the previous section provides a first heuristic estimate of the retrieval threshold in the presence of biased patterns. We now show that the same bias dependent correction can be recovered analytically by means of \textit{Guerra's interpolation method} \cite{guerra_broken}, within the RS framework.

\par\medskip
Our goal is to derive the quenched statistical pressure $\mathcal{A}^{(P)}(\beta,g)$ \eqref{eq:stat-pres-Definition} in the high-storage regime. Since we are interested in the retrieval region, we isolate, without loss of generality, the first of the stored patterns, as the candidate retrieved pattern, and separate its contribution from that of the remaining $K-1$ patterns. The latter act as a source of cross-talk noise. Accordingly, the partition function can be rewritten as
\begin{equation}
\begin{array}{lll}
Z_{g,N}^{(P)}(\beta\,|\,\bm{\xi})
=\SOMMA{\bm{\sigma}\in\Omega}{}
&\exp\left[\dfrac{\beta N }{2}\big(\tilde{m}_1(\bm\sigma|\bm\xi)\big)^P
-\dfrac{\beta g}{2}N\left(M(\bm\sigma) - b\right)^2\right.
\\\\
&\left.
+\dfrac{\beta}{N^{P-1}}\SOMMA{\mu>1}{K}\sum_{(i_1, \cdots, i_P)=1}^{N}
\eta_{i_1}^\mu\cdots\eta_{i_P}^\mu\,\sigma_{i_1}\cdots\sigma_{i_P}\right].
\end{array}
\label{partition-function_0}
\end{equation}
The first term represents the signal associated with the candidate retrieved pattern $\bm \xi^1$, while the last term collects the cross-talk noise generated by all the non-retrieved patterns $\bm \xi^{\mu>1}$.

\par\medskip
In order to implement Guerra's interpolation scheme, we approximate the contribution of the non-retrieved patterns by means of the CLT. In the high-storage regime, the sum over $\mu>1$ gives rise to an effective Gaussian random term, so that the partition function can be recast as
\begin{equation}
\begin{array}{lll}
Z_{g,N}^{(P)}(\beta\,|\,\bm{\xi^1}, \bm \lambda)
=\SOMMA{\bm{\sigma}\in\Omega}{}
&\exp\left[\dfrac{\beta N }{2}\big(\tilde{m}_1(\bm\sigma|\bm\xi)\big)^P
-\dfrac{\beta g}{2}N\left(M(\bm\sigma) - b\right)^2\right.
\\\\
&\left.
+\dfrac{\beta \sqrt{K}}{2 N^{P-1}}\SOMMA{i_1, \cdots, i_P=1}{N}
\lambda_{i_1\cdots i_P}\,\sigma_{i_1}\cdots\sigma_{i_P}\right],
\end{array}
\label{partition-function_start}
\end{equation}
where $\bm \lambda = \{\lambda_{i_1, \hdots, i_P}\}_{i_1, \hdots, i_P =1, \hdots, N}$ and $\lambda_{i_1\cdots i_P}\sim\mathcal{N}(0,1)$ i.i.d. for $i_1, \hdots, i_P =1, \hdots, N$.

Following \cite{guerra_broken}, we introduce an interpolating parameter $t\in[0,1]$ and define the interpolating partition function as
\begin{equation}
\begin{aligned}
Z_{g,N}^{(P)}(\beta \vert \bm{\xi^1}, \bm \lambda, \bm J; t)
&=
\sum_{\bm\sigma\in\Omega}\,
\exp\Bigg[
t \frac{\beta N }{2}(\tilde{m}_1(\bm \sigma \vert \bm \xi))^P
+\beta g N b M(\bm \sigma)
-\dfrac{\beta g}{2}N t (M(\bm \sigma))^2
-\dfrac{\beta g}{2}N b^2
\\
&\qquad\qquad
+\sqrt{t}\dfrac{\beta \sqrt{K}}{2 N^{P-1}}\SOMMA{i_1, \cdots, i_P=1}{N}
\lambda_{i_1\cdots i_P}\,\sigma_{i_1}\cdots\sigma_{i_P}
\\
&\qquad\qquad
+(1-t)\psi N \tilde{m}_1(\bm \sigma \vert \bm \xi)
+(1-t)\tilde\psi N M(\bm \sigma)
+\sqrt{1-t}\,A\sum_i J_i \sigma_i
\Bigg],
\end{aligned}
\label{eq:interpolating_Z}
\end{equation}
where $t\in[0,1]$, the constants $\psi$, $\tilde\psi$, and $A$ will be fixed \textit{a posteriori}, and $\{J_i\}_{i=1,\dots,N}$ are i.i.d.\ standard Gaussian random variables.

We then introduce the finite-volume interpolating quenched statistical pressure
\begin{equation}
\mathcal{A}_{N}^{(P)}(\beta,g;t)
=
\frac{1}{N}\mathbb{E}_{\bm{\xi}^1, \bm \lambda, \bm J}\ln Z_{g,N}^{(P)}(\beta\,|\,\bm{\xi^1}, \bm \lambda, \bm J;t),
\label{eq:stat-pres-interpolating}
\end{equation}
and denote by
\begin{equation}
\mathcal{A}^{(P)}(\beta,g;t)
=
\lim_{N\to\infty}\mathcal{A}_{N}^{(P)}(\beta,g;t)
\end{equation}
its thermodynamic limit. 

By construction, at $t=1$ we recover the original interacting model, so that
\begin{equation}
\mathcal{A}^{(P)}(\beta,g;t=1)=\dfrac{1}{N} \mathbb{E}_{\bm \xi^1, \bm \lambda} \log Z_{g,N}^{(P)}(\beta \vert \bm \xi^1, \bm \lambda, \bm J; t=1)=\mathcal{A}^{(P)}(\beta,g).
\end{equation}
At $t=0$, on the other hand, the model reduces to an effective one-body system, which is explicitly solvable: the neurons decouple and interact only with suitably chosen external fields that reproduce the relevant low-order statistics of the original interacting problem.

From now on, whenever no confusion arises, we omit the explicit dependence of the order parameters on $\bm\sigma$ and $\bm\xi$.

The relation between the two endpoints of the interpolation follows from the Fundamental Theorem of Calculus:
\begin{equation}
\label{eq:T_o_C}
\mathcal{A}^{(P)}(\beta,g;t=1)
=
\mathcal{A}^{(P)}(\beta,g;t=0)
+\int_0^1 dt\,\frac{\partial \mathcal{A}^{(P)}(\beta,g;s)}{\partial s}\Big|_{s=t}.
\end{equation}
We now evaluate the two contributions separately.

\paragraph{One-body contribution.}
The first term is readily computed. Setting $t=0$ in \eqref{eq:interpolating_Z}, we obtain
\begin{equation}
\begin{aligned}
Z_{g,N}^{(P)}(\beta \vert \bm{\xi^1}, \bm \lambda, \bm J; t=0)
&=
\sum_{\bm\sigma\in\Omega}\,
\exp\Bigg[
\beta g b N M
-\dfrac{\beta g}{2}N b^2
+\psi N \tilde{m}_1
+\tilde\psi N M
+A\sum_i J_i \sigma_i
\Bigg].
\end{aligned}
\label{eq:interpolating_Z_ONE}
\end{equation}
At this stage the spins decouple, and the partition function factorizes into a product of one site contributions. Performing the Gaussian average and using standard algebraic manipulations, one finds
\begin{equation}
\begin{aligned}
Z_{g,N}^{(P)}(\beta \vert \bm{\xi^1}, \bm J; t=0)
&=
\exp\Bigg[
-\dfrac{\beta g}{2}N b^2
\Bigg]
\,2^N
\cosh^N\Bigg[
\beta g b + \psi \eta^1 + \tilde\psi + A J
\Bigg],
\end{aligned}
\label{eq:interpolating_Z_one_end}
\end{equation}
and therefore
\begin{equation}
\begin{aligned}
\mathcal{A}^{(P)}(\beta,g;t=0)
&=
\log 2
-\dfrac{\beta g b^2}{2}
+\mathbb{E}_{x,\eta^1}\log\cosh\Bigg[
\psi \eta^1 + \beta g b + \tilde\psi + x\sqrt{A^2}
\Bigg].
\end{aligned}
\label{eq:one_body}
\end{equation}
The details of this computation are reported in Appendix~\ref{sec:proofs}.

\paragraph{Streaming term.}
At finite volume, the derivative of the interpolating quenched pressure with respect to $t$ is computed in Appendix~\ref{sec:proofs} and reads
\begin{equation}
\begin{aligned}
\partial_t \mathcal{A}_{N}^{(P)}(\beta,g; t)
&= \frac{\beta}{2}\Bigg[
\langle (\tilde m_1)^P\rangle
-\frac{2\psi}{\beta}\langle m_1\rangle
\Bigg]
-\frac{\beta g }{2}\Bigg[
\langle M^2\rangle
+\frac{2\tilde\psi}{\beta g }\langle M\rangle
\Bigg]
\\
&\quad
+\frac{\beta^2 K}{8N^{P-1}}\Big[
1-\langle q_{12}^P\rangle
\Big]
-\frac{A^2}{2}\Big[
1-\langle q_{12}\rangle
\Big].
\end{aligned}
\label{eq:dtf}
\end{equation}

We now impose the RS ansatz, namely that the relevant order parameters self-average around deterministic values in the thermodynamic limit. More precisely, for any order parameter $X$ with limiting value $\bar X$, we assume
\begin{equation}
\lim_{N\to\infty}\mathbb{P}_N(X)=\delta(X-\bar X).
\label{eq:RSassumption}
\end{equation}
Under this assumption, the streaming term simplifies in the thermodynamic limit to
\begin{equation}
\begin{aligned}
\partial_t \mathcal{A}^{(P)}(\beta,g; t)
=
-\frac{\beta(P-1)}{2}\bar{\tilde m}^{P}
+\frac{\beta g }{2}\bar{M}^{2}
-\frac{\beta^2\gamma}{8}\left[1 - P\bar{q}^{P-1} + (P-1)\bar q^{P}\right],
\end{aligned}
\label{eq:dtff}
\end{equation}
provided that we have used the definition of the load \eqref{eq:load} and the interpolating parameters are chosen as
\begin{equation}
\begin{aligned}
\psi &= \beta\dfrac{P}{2}\bar{\tilde m}^{P-1},
\qquad
\tilde\psi = -\beta g \bar M,
\qquad
A^2 = \dfrac{\beta^2 \gamma P}{2}\bar{q}^{P-1}.
\label{eq:constants}
\end{aligned}
\end{equation}

\paragraph{RS statistical pressure and self consistency equations.}
Substituting \eqref{eq:one_body} and \eqref{eq:dtff} into \eqref{eq:T_o_C}, we obtain the RS expression for the quenched statistical pressure in the thermodynamic limit:
\begin{equation}
\begin{aligned}
\mathcal{A}^{(P)}(\beta,g)
&=
\log 2
+\mathbb{E}_{x,\eta^1}\log\cosh\Bigg[
\beta \dfrac{P}{2}\bar{\tilde m}^{P-1}\eta^1
+\beta g (b-\bar M)
+x\beta\sqrt{\dfrac{\gamma P}{2}\bar q^{P-1}}
\Bigg]
\\
&\quad
-\frac{\beta(P-1)}{2}\bar{\tilde m}^{P}
-\frac{\beta g }{2}(b^2-\bar M^2)
-\frac{\beta^2\gamma}{8}\left[1 - P\bar{q}^{P-1} + (P-1)\bar q^{P}\right].
\end{aligned}
\label{eq:A_RS}
\end{equation}

The corresponding order parameters satisfy the following self-consistency equations
\begin{equation}
\label{eq:self_n}
\begin{aligned}
\bar{\tilde{m}}
&=
\mathbb{E}_{x,\xi^1}\Bigg\{
\big(\xi^1-b\big)
\tanh\Bigg[
\beta \dfrac{P}{2}\bar{\tilde m}^{P-1}\big(\xi^1-b\big)
+\beta g (b-\bar{M})
+x\beta\sqrt{\dfrac{\gamma P}{2}(1-b^2)^{P}\,\bar{q}^{P-1}}
\Bigg]
\Bigg\},
\\
\bar{q}
&=
\mathbb{E}_{x,\xi^1}\Bigg\{
\tanh^2\Bigg[
\beta \dfrac{P}{2}\bar{\tilde m}^{P-1}\big(\xi^1-b\big)
+\beta g (b-\bar{M})
+x\beta\sqrt{\dfrac{\gamma P}{2}(1-b^2)^{P}\,\bar{q}^{P-1}}
\Bigg]
\Bigg\},
\\
\bar{M}
&=
\mathbb{E}_{x,\xi^1}\Bigg\{
\tanh\Bigg[
\beta \dfrac{P}{2}\bar{\tilde m}^{P-1}\big(\xi^1-b\big)
+\beta g (b-\bar{M})
+x\beta\sqrt{\dfrac{\gamma P}{2}(1-b^2)^{P}\,\bar{q}^{P-1}}
\Bigg]
\Bigg\}.
\end{aligned}
\end{equation}
We stress that here we have performed the rescalings $\bar{\tilde{m}}\to\bar{\tilde{m}}/\sqrt{1-b^2}$, $g(1-b^2)\to g$, and $\beta/(1-b^2)^{P/2}\to\beta$.

\par\medskip
At this stage, it is worth emphasizing the full consistency between the heuristic signal-to-noise argument and the analytical Guerra interpolation scheme. Indeed, the variance of the effective Gaussian noise entering the self-consistency equations is renormalized by the same multiplicative factor $(1-b^2)^P$ already predicted by the signal-to-noise estimate in \eqref{eq:load_bias_corr}. Therefore, the bias does not modify the superlinear scaling of the storage capacity, but reduces its prefactor in a quantitatively consistent way across the two approaches.

\par\medskip
Finally, one can also analyze the zero-temperature limit of the previous equations which can be further simplify to
\begin{equation}
\label{eq:self_n_T_zero}
\begin{aligned}
\bar{\tilde{m}}
&=
\dfrac{1-b^2}{2}\left\{
\mathrm{erf}\left[
\dfrac{\dfrac{P}{2}\bar{\tilde m}^{P-1}\big(1-b\big) + g (b-\bar{M})}
{\sqrt{\gamma P(1-b^2)^{P}}}
\right]
+
\mathrm{erf}\left[
\dfrac{\dfrac{P}{2}\bar{\tilde m}^{P-1}\big(1+b\big) - g (b-\bar{M})}
{\sqrt{\gamma P(1-b^2)^{P}}}
\right]
\right\},
\\
\bar{M}
&=
\dfrac{1+b}{2}\mathrm{erf}\left[
\dfrac{\dfrac{P}{2}\bar{\tilde m}^{P-1}\big(1-b\big) + g (b-\bar{M})}
{\sqrt{\gamma P(1-b^2)^{P}}}
\right]
-\dfrac{1-b}{2}\mathrm{erf}\left[
\dfrac{\dfrac{P}{2}\bar{\tilde m}^{P-1}\big(1+b\big) - g (b-\bar{M})}
{\sqrt{\gamma P(1-b^2)^{P}}}
\right].
\end{aligned}
\end{equation}
These equations will be used to investigate the phase diagram and the retrieval performance of the model (see Appendix \ref{app:Tnullo} for the calculation details).

\begin{figure}[t]
    \centering
    \includegraphics[width=15cm]{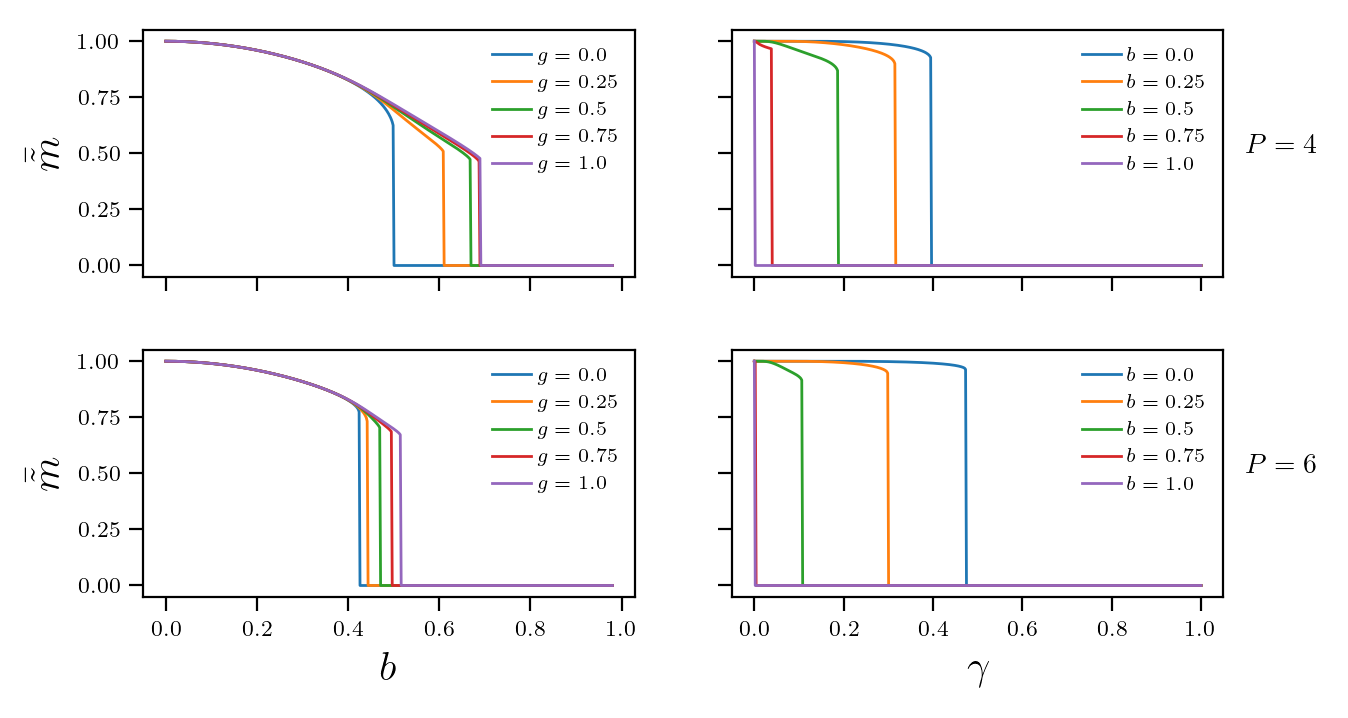}
    \caption{Numerical resolution of zero temperature self consistency equation is presented. The upper row corresponds to $P=4$, and the lower row to $P=6$. In the first column, $\gamma=0.1$ is held fixed, while different values of $b$ and $g$ are explored. In the second column, $g=1.0$ is fixed, and both the load $\gamma$ and bias $b$ are varied.}
    \label{fig:zero_temp_new}
\end{figure}

\begin{figure}[t]
    \centering
    \includegraphics[width=15cm]{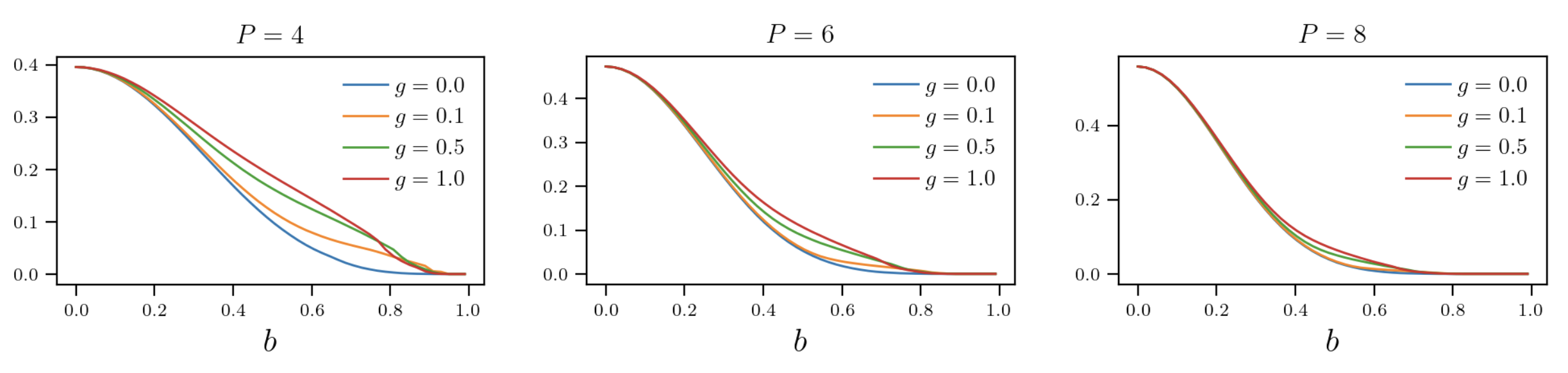}
    \caption{Behaviour of the critical load of the network at null temperature $T =0$ for different values of $g$ and $b$. In each panel we change the grade of interactions of the network (from left to right, $P=4,6,8$) and we analyze different values of $g$ from $g=0$, when the term tuning the constraint on the activity of the network vanishes, to $g=1$. We can see that $g$ plays the role of a light booster for the critical load, whose contribution decreases when the degree of interactions increases. Moreover, the critical load becomes smaller as the bias of the patterns increases.}
    \label{fig:diff_P}
\end{figure}

\par\medskip
To corroborate our results, we provide a numerical analysis. 
First, in Fig.~\ref{fig:zero_temp_new} we plot the self consistency equations \eqref{eq:self_n_T_zero}. Then, in Fig.~\ref{fig:diff_P}, we use these equations to show how the critical load of the network depends on the bias parameters \(b\) and \(g\) for different interaction orders \(P\).

\begin{figure}[t]
    \centering
    \includegraphics[width=15cm]{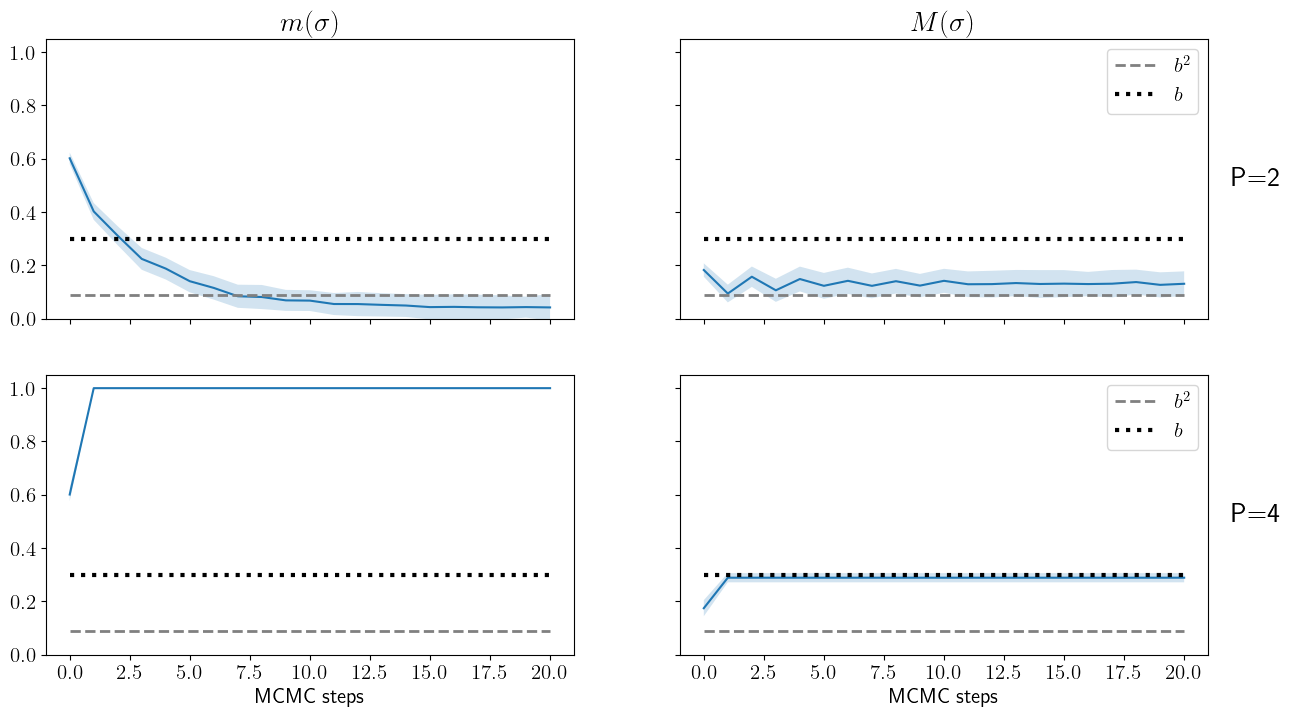}
    \caption{MCMC simulation, for $N=1000$, $K=O(N^2)$, $b=0.3$ and $g=1.0$, of the Mattis magnetization $m(\sigma)$ and the mean activity $M(\sigma)$ with respect to different values of the degree of interactions $P$. Starting from a configuration which has an overlap with $\xi^1$ equal to $0.6$, the standard Hopfield case $(P=2)$ is not able to retrieve the pattern, whereas the dense associative memory with $P=4$ does this and, moreover, the mean activity stabilizes around the value of the bias $b$, as expected.}
    \label{fig:dens_VS_pair1}
\end{figure}

\begin{figure}[t]
    \centering
    \includegraphics[width=15cm]{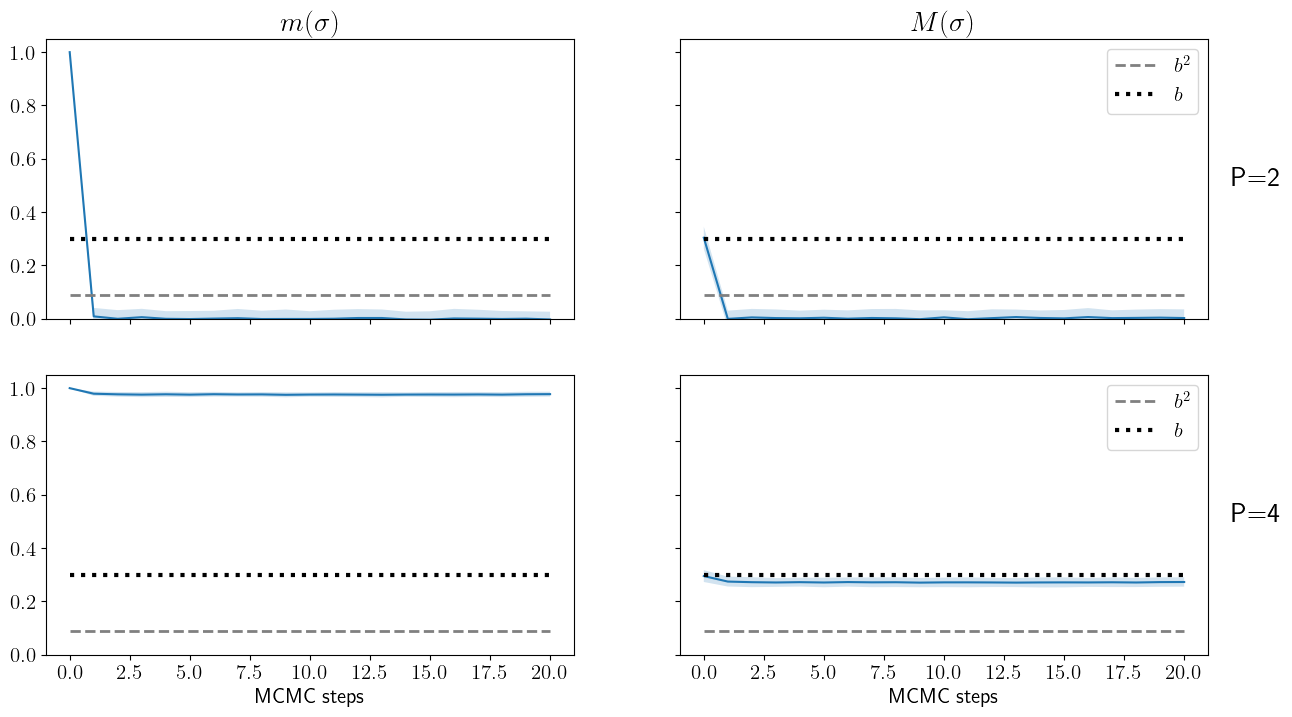}
    \caption{MCMC simulation for $N=1000$, $K= 0.05 N^3$, $b=0.3$, $g=1.0$. In the dense, low-load limit, the system successfully retrieves the stored pattern even in the presence of bias, yielding both $m(\sigma)=1$ and $M(\sigma)=b$. Conversely, for $P=2$ retrieval fails, as the load exceeds the network storage capacity. In all cases, the dynamics are initialized from a network configuration such that $\bm{\sigma}^{(0)} \cdot \bm{\xi}^1 \approx 1.0$. For $P=2$, the trajectory rapidly departs from the initial condition, whereas for $P=4$ the system remains trapped in the initial configuration, indicating stability.}
    \label{fig:dens_VS_pair2}
\end{figure}

In the next two figures (Figs.~\ref{fig:dens_VS_pair1}--\ref{fig:dens_VS_pair2}), we compare, through Markov Chain Monte Carlo (MCMC) simulations, the retrieval performance of the biased pairwise Hopfield model with that of a dense biased associative memory with interaction order \(P=4\). We stress that the stored patterns are biased: specifically, the correlation between patterns is at least \(b^2\), which corresponds to the random case baseline.

In Fig.~\ref{fig:dens_VS_pair1}, we initialize the MCMC dynamics from a state \(\bm{\sigma}^{(0)}\) whose Mattis magnetization with respect to the first pattern is equal to \(0.6\), and we monitor the evolution of the Mattis magnetization \(m_1(\sigma)\) and of the mean activity \(M(\sigma)\) in the regime where the number of patterns scales as \(K = O(N^2)\). The bias is fixed at \(b=0.3\) and the constraint at \(g=1.0\). The standard pairwise Hopfield model is unable to retrieve the patterns: indeed, after a few update steps, the magnetization decays to zero. By contrast, for \(P=4\) the pattern is successfully retrieved, as shown by the thermalized value of the Mattis magnetization, which approaches \(1.0\), while the mean activity stabilizes around \(b\), consistently with the average activity of the biased patterns.

In Fig.~\ref{fig:dens_VS_pair2}, we further perform MCMC simulations close to the saturation load in order to test the stability of the biased patterns in a more challenging regime. We set \(K = O(N^3)\), with \(b=0.3\), \(g=1.0\), and \(\bm{\sigma}^{(0)}=\bm{\xi}^1\), and we let the system evolve while tracking both the Mattis magnetization with respect to the first pattern and the mean activity of the network. As can be seen, in the Hopfield model with \(P=2\) (top row) the stability of the pattern quickly deteriorates, since both the mean activity and the magnetization with respect to the first pattern decay to zero within a few steps. In contrast, the dense network with \(P=4\) (bottom row) continues to perform well, keeping both the mean activity and the Mattis magnetization of the first pattern close to \(b\) and \(1.0\), respectively.

\section{Conclusion}

In this work, we investigated the retrieval properties of dense associative memories in the presence of biased patterns. The model we considered extends the standard $P-$spin Hopfield model architecture \cite{HopKro1} to the case in which the stored patterns have non-zero mean activity. In this setting, the usual Hebbian prescription must be modified by recentering and rescaling the pattern entries, and a further, following \cite{amit1987information}, soft constraint on the global activity must be introduced in order to ensure that the network evolves toward configurations compatible with the average activation of the stored memories.

\par\medskip
Our analysis combined two complementary perspectives. On the one hand, the signal-to-noise argument provided a simple and transparent zero-temperature estimate of the retrieval threshold. This heuristic computation showed that the effect of the bias is to reduce the effective storage capacity by a multiplicative factor $(1-b^2)^P$, while leaving unchanged the superlinear scaling $K\sim N^{P-1}$ that characterizes dense higher-order networks. On the other hand, the thermodynamic analysis based on Guerra's interpolation method allowed us to derive the quenched free energy within the RS framework and to obtain the corresponding self consistency equations for the relevant order parameters.

A central outcome of the paper is the full consistency between these two approaches. Indeed, the same bias dependent renormalization predicted by the signal-to-noise analysis emerges analytically in the variance of the effective cross-talk noise entering the RS equations. This agreement provides a clear physical interpretation of the role of bias: it does not alter the qualitative capacity scaling of the model, but quantitatively suppresses retrieval by increasing the effective difficulty of separating signal from noise.

More broadly, our results show that dense associative memories remain robust in the high storage regime even when the stored information is statistically unbalanced, provided that the learning rule and the network dynamics are modified consistently. The interplay among pattern bias, storage load, thermal noise, and activity constraint gives rise to a rich phase structure, whose numerical exploration can clarify the computational regions in which retrieval remains possible.

\par\medskip
Several directions deserve further investigation. A first natural extension concerns the study of Replica Symmetry Breaking effects \cite{MPV}, which may become relevant in parts of the phase diagram where glassy behavior competes with retrieval. A second interesting problem is the characterization of the dynamical properties of the model beyond one step stability arguments, in particular the size of the basins of attraction and the convergence time toward biased memories. Another interesting extension of the present framework concerns hetero-associative architectures, where different networks interact and exchange information, potentially enhancing retrieval and generalization capabilities \cite{agliari2025generalized}. Finally, it would be worth exploring whether analogous bias induced renormalizations persist in sparse or diluted higher-order architectures, as well as in modern Hopfield networks arising in machine learning \cite{lucibello2024exponential, albanese2025yet}.

\acknowledgments
L.A. acknowledges funding from the project “Patto Territoriale del Sistema Universitario Pugliese” (CUP F61B23000370006). \\ 
L.A. acknowledge the PRIN 2022 grant {\em Statistical Mechanics of Learning Machines} number 20229T9EAT funded by the Italian Ministry of University and Research (MUR) in the framework of European Union - Next Generation EU.\\
L.A. acknowledges funding also by the PRIN 2022 grant {\em “Stochastic Modeling of Compound Events (SLIDE)”} n. P2022KZJTZ funded by the Italian Ministry of University and Research (MUR) in the framework of European Union - Next Generation EU.
\newline
A.A. acknowledges BULBUL “Brain-inspired ULtra-Fast \& Ultra-sharp neural networks” for support via post-Lauream research fellowships 
“Statistical mechanics of hetero-associative neural networks” (CUP F85F21006230001, D.D. n. 325/29-09-2025).
\newline
L.A. and A.A. are members of the GNFM group of INdAM which is acknowledged too.


\begin{thebibliography}{10}

\bibitem{super}
E.~Agliari, L.~Albanese, F.~Alemanno, A.~Alessandrelli, A.~Barra, F.~Giannotti, D.~Lotito, and D.~Pedreschi.
\newblock Dense hebbian neural networks: a replica symmetric picture of supervised learning.
\newblock {\em Physica A: Statistical Mechanics and its Applications}, 626:129076, 2023.

\bibitem{unsup}
E.~Agliari, L.~Albanese, F.~Alemanno, A.~Alessandrelli, A.~Barra, F.~Giannotti, D.~Lotito, and D.~Pedreschi.
\newblock Dense hebbian neural networks: A replica symmetric picture of unsupervised learning.
\newblock {\em Physica A: Statistical Mechanics and its Applications}, 627:129143, 2023.

\bibitem{agliari2019dreaming}
E.~Agliari, F.~Alemanno, A.~Barra, and A.~Fachechi.
\newblock Dreaming neural networks: rigorous results.
\newblock {\em Journal of Statistical Mechanics: Theory and Experiment}, 2019(8):083503, 2019.

\bibitem{agliari2025generalized}
E.~Agliari, A.~Alessandrelli, A.~Barra, M.~S. Centonze, and F.~Ricci-Tersenghi.
\newblock Generalized hetero-associative neural networks.
\newblock {\em Journal of Statistical Mechanics: Theory and Experiment}, 2025(1):013302, 2025.

\bibitem{agliari2020tolerance}
E.~Agliari and G.~De~Marzo.
\newblock Tolerance versus synaptic noise in dense associative memories.
\newblock {\em The European Physical Journal Plus}, 135(11):1--22, 2020.

\bibitem{agliari2022storing}
E.~Agliari, F.~E. Leonelli, and C.~Marullo.
\newblock Storing, learning and retrieving biased patterns.
\newblock {\em Applied Mathematics and Computation}, 415:126716, 2022.

\bibitem{Albanese2021}
L.~Albanese, F.~Alemanno, A.~Alessandrelli, and A.~Barra.
\newblock Replica symmetry breaking in dense hebbian neural networks.
\newblock {\em Journal of Statistical Physics}, 189(2):1--41, 2022.

\bibitem{albanese2024replica}
L.~Albanese, A.~Alessandrelli, A.~Annibale, and A.~Barra.
\newblock Replica symmetry breaking in supervised and unsupervised hebbian networks.
\newblock {\em Journal of Physics A: Mathematical and Theoretical}, 57(16):165003, 2024.

\bibitem{albanese2025yet}
L.~Albanese, A.~Alessandrelli, A.~Barra, and P.~Sollich.
\newblock Yet another exponential {H}opfield model.
\newblock {\em Physica A: Statistical Mechanics and its Applications}, page 131223, 2025.

\bibitem{Amit}
D.~J. Amit.
\newblock {\em Modeling brain function: The world of attractor neural networks}.
\newblock Cambridge university press, 1989.

\bibitem{AGS}
D.~J. Amit, H.~Gutfreund, and H.~Sompolinsky.
\newblock Storing infinite numbers of patterns in a spin-glass model of neural networks.
\newblock {\em Physical Review Letters}, 55:1530--1533, 1985.

\bibitem{amit1987information}
D.~J. Amit, H.~Gutfreund, and H.~Sompolinsky.
\newblock Information storage in neural networks with low levels of activity.
\newblock {\em Physical Review A}, 35(5):2293, 1987.

\bibitem{Baldi}
P.~Baldi and S.~S. Venkatesh.
\newblock Number of stable points for spin-glasses and neural networks of higher orders.
\newblock {\em Physical Review Letters}, 58, 1987.

\bibitem{Bovier}
A.~Bovier and B.~Niederhauser.
\newblock The spin-glass phase-transition in the {H}opfield model with p-spin interactions.
\newblock {\em Advances in Theoretical and Mathematical Physics}, 5:1001--1046, 8 2001.

\bibitem{fachechi2019dreaming}
A.~Fachechi, E.~Agliari, and A.~Barra.
\newblock Dreaming neural networks: forgetting spurious memories and reinforcing pure ones.
\newblock {\em Neural Networks}, 112:24--40, 2019.

\bibitem{Gardner}
E.~Gardner.
\newblock Multiconnected neural network models.
\newblock {\em Journal of Physics A: General Physics}, 20, 1987.

\bibitem{guerra_broken}
F.~Guerra.
\newblock Broken replica symmetry bounds in the mean field spin glass model.
\newblock {\em Communications in Mathematical Physics}, 233:1--12, 2003.

\bibitem{Hopfield}
J.~J. Hopfield.
\newblock Neural networks and physical systems with emergent collective computational abilities.
\newblock {\em Proceedings of the National Academy of Sciences of the United States of America}, 79:2554--2558, 1982.

\bibitem{Krotov2018}
D.~Krotov and J.~Hopfield.
\newblock Dense associative memory is robust to adversarial inputs.
\newblock {\em Neural Computation}, 30:3151--3167, 2018.

\bibitem{HopKro1}
D.~Krotov and J.~J. Hopfield.
\newblock Dense associative memory for pattern recognition.
\newblock {\em Advances in Neural Information Processing Systems}, pages 1180--1188, 2016.

\bibitem{lowe1999storage}
M.~Lowe.
\newblock On the storage capacity of the {H}opfield model with biased patterns.
\newblock {\em IEEE Transactions on Information Theory}, 45(1):314--318, 1999.

\bibitem{lucibello2024exponential}
C.~Lucibello and M.~M{\'e}zard.
\newblock Exponential capacity of dense associative memories.
\newblock {\em Physical Review Letters}, 132(7):077301, 2024.

\bibitem{MPV}
M.~Mézard, G.~Parisi, and M.~A. Virasoro.
\newblock {\em Spin glass theory and beyond: An Introduction to the Replica Method and Its Applications}, volume~9.
\newblock World Scientific Publishing Company, 1987.

\bibitem{nishimori2001statistical}
H.~Nishimori.
\newblock {\em Statistical physics of spin glasses and information processing: an introduction}.
\newblock Number 111. Clarendon Press, 2001.

\bibitem{peretto1984collective}
P.~Peretto.
\newblock Collective properties of neural networks: a statistical physics approach.
\newblock {\em Biological cybernetics}, 50(1):51--62, 1984.

\bibitem{ramsauer2020hopfield}
H.~Ramsauer, B.~Sch{\"a}fl, J.~Lehner, P.~Seidl, M.~Widrich, T.~Adler, L.~Gruber, M.~Holzleitner, M.~Pavlovi{\'c}, G.~K. Sandve, et~al.
\newblock {H}opfield networks is all you need.
\newblock {\em arXiv preprint arXiv:2008.02217}, 2020.

\bibitem{tsodyks1988enhanced}
M.~V. Tsodyks and M.~V. Feigel'man.
\newblock The enhanced storage capacity in neural networks with low activity level.
\newblock {\em EPL (Europhysics Letters)}, 6(2):101--105, 1988.

\end{thebibliography}

\appendix
\section{Computation of $t-$derivatives and one body terms of the quenched statistical pressure}
\label{sec:proofs}
Let us start from the computation of the derivative with respect to $t$ of the quenched statistical pressure \eqref{eq:stat-pres-interpolating}.

\begin{align}
    d_t \mathcal{A}_N(\beta, g; t)&= \dfrac{1}{N} \mathbb{E}_{\bm \xi^1, \bm \lambda} \dfrac{1}{Z_{g,N}(\beta \vert \bm \xi^1, \bm \lambda, \bm J; t)} \sum_{\bm \sigma} B_{g,N}(\beta \vert \bm \xi^1, \bm \lambda, \bm J; t) \left[ \frac{\beta N }{2}(\tilde{m}_1(\bm \sigma \vert \bm \xi))^P
\right. \notag \\
&-\dfrac{\beta g}{2}N (M(\bm \sigma))^2 -\tilde\psi N M(\bm \sigma) -\,\dfrac{A}{2\sqrt{1-t}}\sum_i J_i \sigma_i\notag\\
&\left.+\dfrac{\beta \sqrt{K}}{2 \sqrt{t} N^{P-1}}\SOMMA{i_1, \cdots, i_P=1}{N} 
\lambda_{i_1\cdots i_P}\,\sigma_{i_1}\cdots\sigma_{i_P} -\psi N \tilde{m}_1(\bm \sigma \vert \bm \xi)
\right]
\end{align}
where $B_{g,N}(\beta \vert \bm \xi^1, \bm \lambda, \bm J; t)$ is the Boltzmann weight, namely 
\begin{align}
    B_{g,N}(\beta \vert \bm \xi^1, \bm \lambda, \bm J; t) =& \exp\Bigg[
t \frac{\beta N }{2}(\tilde{m}_1(\bm \sigma \vert \bm \xi))^P
+\beta g N b M(\bm \sigma)
-\dfrac{\beta g}{2}N t (M(\bm \sigma))^2
-\dfrac{\beta g}{2}N b^2
\\
&\qquad\qquad
+\sqrt{t}\dfrac{\beta \sqrt{K}}{2 N^{P-1}}\SOMMA{i_1, \cdots, i_P=1}{N}
\lambda_{i_1\cdots i_P}\,\sigma_{i_1}\cdots\sigma_{i_P}
\\
&\qquad\qquad
+(1-t)\psi N \tilde{m}_1(\bm \sigma \vert \bm \xi)
+(1-t)\tilde\psi N M(\bm \sigma)
+\sqrt{1-t}\,A\sum_i J_i \sigma_i
\Bigg].
\end{align}

We can apply the definition of the quenched average \eqref{eq:totalaver} on the trivial terms in order to get 
\begin{align}
    &d_t \mathcal{A}_N(\beta, g; t)= \frac{\beta}{2}\l (\tilde{m}_1(\bm \sigma \vert \bm \xi))^P \r
-\dfrac{\beta g}{2} \l (M(\bm \sigma))^2 \r -\tilde\psi \l M(\bm \sigma) \r - \psi \l \tilde m_1(\bm \sigma \vert \bm \xi) \r \notag \\
&+\dfrac{1}{N} \mathbb{E}_{\bm \xi^1, \bm \lambda} \dfrac{1}{Z_{g,N}(\beta \vert\bm \xi^1, \bm \lambda, \bm J; t)} \sum_{\bm \sigma} B_{g,N}(\beta \vert \bm \xi^1, \bm \lambda, \bm J; t)\dfrac{\beta \sqrt{K}}{2 \sqrt{t} N^{P-1}}\SOMMA{i_1, \cdots, i_P=1}{N}
\lambda_{i_1\cdots i_P}\,\sigma_{i_1}\cdots\sigma_{i_P} 
\notag \\
&-\dfrac{1}{N} \mathbb{E}_{\bm \xi^1, \bm \lambda} \dfrac{1}{Z_{g,N}(\beta \vert\bm \xi^1, \bm \lambda, \bm J; t)} \sum_{\bm \sigma} B_{g,N}(\beta \vert \bm \xi^1, \bm \lambda, \bm J; t) 
\,\dfrac{A}{2\sqrt{1-t}}\sum_i J_i \sigma_i.
\label{eq:appdt}
\end{align}

In order to compute the last two terms, we need to apply the \textit{Stein's lemma} (also known as \textit{Wick's theorem}) which states that for any smooth function $f(\bm J)$ of a centered unit Gaussian variable $\bm J$ for which the two expectations $\mathbb{E}\left( \bm J f(\bm J)\right)$ and $\mathbb{E}\left( \partial_{\bm J} f(\bm J)\right)$ both exist the relation
\begin{align}
\mathbb{E}_{\bm J}( \bm J f(\bm J) ) = \mathbb{E}_{\bm J} \Bigg( \frac{\partial}{\partial \bm J} f(\bm J) \Bigg),
\label{eq:steins_app}
\end{align}
holds. Since the last two terms of \eqref{eq:appdt} have analogous computations, we will show the computations only for the first one. 
\begin{align}
    & \dfrac{\beta \sqrt{K}}{2 \sqrt{t} N^{P}} \SOMMA{i_1, \cdots, i_P=1}{N} \mathbb{E}_{\bm \xi^1, \bm \lambda} \left[\dfrac{1}{Z_{g,N}(\beta \vert \bm \xi^1, \bm \lambda, \bm J; t)} \sum_{\bm \sigma} B_{g,N}(\beta \vert \bm \xi^1, \bm \lambda, \bm J; t)
\lambda_{i_1\cdots i_P}\,\sigma_{i_1}\cdots\sigma_{i_P}  \right]\notag \\
&= \dfrac{\beta \sqrt{K}}{2 \sqrt{t} N^{P}} \SOMMA{i_1, \cdots, i_P=1}{N} \mathbb{E}_{\bm \xi^1, \bm \lambda} \left[\partial_{\lambda_{i_1\cdots i_P}} \left(\dfrac{1}{Z_{g,N}(\beta \vert \bm \xi^1, \bm \lambda, \bm J; t)} \sum_{\bm \sigma} B_{g,N}(\beta \vert \bm \xi^1, \bm \lambda, \bm J; t)\sigma_{i_1}\cdots\sigma_{i_P} \right)\right]; \\
&\partial_{\lambda_{i_1\cdots i_P}} \left(\dfrac{1}{Z_{g,N}(\beta \vert \bm \xi^1, \bm \lambda, \bm J; t)} \sum_{\bm \sigma} B_{g,N}(\beta \vert \bm \xi^1, \bm \lambda, \bm J; t) \sigma_{i_1}\cdots\sigma_{i_P}\right) = \notag \\
&\hspace{3cm}\partial_{\lambda_{i_1\cdots i_P}} \left(\dfrac{1}{Z_{g,N}(\beta \vert \bm \xi^1, \bm \lambda, \bm J; t)} \right)\sum_{\bm \sigma} B_{g,N}(\beta \vert \bm \xi^1, \bm \lambda, \bm J; t) \sigma_{i_1}\cdots\sigma_{i_P} \notag \\
&\hspace{3cm}+ \dfrac{1}{Z_{g,N}(\beta \vert \bm \xi^1, \bm \lambda, \bm J; t)} \sum_{\bm \sigma} \partial_{\lambda_{i_1\cdots i_P}} \left(B_{g,N}(\beta \vert \bm \xi^1, \bm \lambda, \bm J; t) \sigma_{i_1}\cdots\sigma_{i_P}\right); \\
&\partial_{\lambda_{i_1\cdots i_P}} \left(\dfrac{1}{Z_{g,N}(\beta \vert \bm \xi^1, \bm \lambda, \bm J; t)} \right) = -\dfrac{\beta \sqrt{K t}}{2 N^{P-1}} \left(\dfrac{1}{Z_{g,N}(\beta \vert \bm \xi^1, \bm \lambda, \bm J; t)} \right)^2 \sum_{\bm \sigma} B_{g,N}(\beta \vert \bm \xi^1, \bm \lambda, \bm J; t) \sigma_{i_1}\cdots\sigma_{i_P}; \\
&\partial_{\lambda_{i_1\cdots i_P}} \left(B_{g,N}(\beta \vert \bm \xi^1, \bm \lambda, \bm J; t) \sigma_{i_1}\cdots\sigma_{i_P}\right) =  \sqrt{t}\dfrac{\beta \sqrt{K}}{2 N^{P-1}} B_{g,N}(\beta \vert \bm \xi^1, \bm \lambda, \bm J; t) \left(\sigma_{i_1}\cdots\sigma_{i_P}\right)^2.
\end{align}
Putting all the terms together we get 
\begin{align}
    &\dfrac{\beta^2 K }{2N^{2P-1}} \SOMMA{i_1, \cdots, i_P=1}{N} \left\{ \mathbb{E}_{\bm \xi^1, \bm \lambda} \left[ \dfrac{1}{Z_{g,N}(\beta \vert \bm \xi^1, \bm \lambda, \bm J; t)} \sum_{\bm \sigma} B_{g,N}(\beta \vert\bm \xi^1, \bm \lambda, \bm J; t) (\sigma_{i_1}\cdots\sigma_{i_P})^2 \right] \right. \notag \\
    &\left. - \mathbb{E}_{\bm \xi}\left(\dfrac{1}{Z_{g,N}(\beta \vert \bm \xi^1, \bm \lambda, \bm J; t)} \sum_{\bm \sigma} B_{g,N}(\beta \vert \bm \xi^1, \bm \lambda, \bm J; t) \sigma_{i_1}\cdots\sigma_{i_P} \right)^2 \right\} \notag \\
    & \dfrac{\beta^2 K }{2N^{2P-1}} \SOMMA{i_1, \cdots, i_P=1}{N} \mathbb{E}_{\bm \xi^1, \bm \lambda} \left[ \omega(\sigma_{i_1}\cdots\sigma_{i_P})^2 - \omega^2 (\sigma_{i_1}\cdots\sigma_{i_P})\right] = \dfrac{\beta^2 K}{2N^{P-1}} \left( 1-\l q_{12}^P \r\right).
    \label{eq:mb_app}
\end{align}
where in the last passage we have applied the definition of the two-replicas overlap \eqref{eq:order_q}. 

In the same way one can compute the last term in \eqref{eq:appdt} getting 
\begin{align}
    -\,\dfrac{A}{2\sqrt{1-t}}\sum_{i=1}^N \mathbb{E}_{\bm \xi^1, \bm \lambda}  \dfrac{1}{Z_{g,N}(\beta \vert \bm \xi^1, \bm \lambda, \bm J; t)} \sum_{\bm \sigma} B_{g,N}(\beta \vert \bm \xi^1, \bm \lambda, \bm J; t) J_i \sigma_i = \dfrac{A^2}{2} \left( 1-\l q_{12} \r \right).
    \label{eq:app_A}
\end{align}

Finally, inserting \eqref{eq:mb_app} and \eqref{eq:app_A} in \eqref{eq:appdt} and applying the definition of the quenched average \eqref{eq:totalaver}, we get the expression of the derivative of the interpolating quenched statistical pressure with respect to $t$ at finite size $N$ \eqref{eq:dtf}, namely
\begin{align}
    \partial_t \mathcal{A}_{N}^{(P)}(\beta,g; t)
&= \frac{\beta}{2}\Bigg[
\langle (\tilde m_1)^P\rangle
-\frac{2\psi}{\beta}\langle m_1\rangle
\Bigg]
-\frac{\beta g }{2}\Bigg[
\langle M^2\rangle
+\frac{2\tilde\psi}{\beta g }\langle M\rangle
\Bigg]
\notag \\
&\quad
+\frac{\beta^2 K}{8N^{P-1}}\Big[
1-\langle q_{12}^P\rangle
\Big]
-\frac{A^2}{2}\Big[
1-\langle q_{12}\rangle
\Big].
\end{align}

In the thermodynamic limit $N \to +\infty$ we apply RS assumption \eqref{eq:RSassumption}. This means that all the centered moments higher than second order vanishes. This allows us to write for any order parameter $x$ whose equilibrium value is $\bar x$
\begin{equation}
 \begin{array}{lll}
     \langle x^P \rangle -P\,\bar{x}^{P-1}\langle x \rangle &=& -(P-1) \bar{x}^{P}+ \SOMMA{k=2}{P} \begin{pmatrix}P\\k\end{pmatrix} \langle (x-\bar{x})^k \rangle \bar{x}^{P-k},
\end{array}  
\label{eq:RS_pq_Potenziali}
\end{equation}
where the last term vanishes in RS assumption. \\
With some algebraic manipulation and putting the constants $\psi, \ \tilde \psi, \ A$ as in \eqref{eq:constants} we get the thermodynamic streaming term. 

As far as one body terms concerns, one can rewrite the interpolating partition function \eqref{eq:interpolating_Z_ONE} at $t=0$ as 
\begin{equation}
\begin{aligned}
Z_{g,N}^{(P)}(\beta \vert \bm \xi^1, \bm \lambda, \bm J; t=0)
&=
\sum_{\bm\sigma\in\Omega}\,
\exp\Bigg[
\beta g b \sum_{i=1}^N \sigma_i
-\dfrac{\beta g}{2}N b^2
+\psi \sum_{i=1}^N \eta_i^1 \sigma_i
+\tilde\psi \sum_{i=1}^N \sigma_i
+A\sum_{i=1}^N J_i \sigma_i
\Bigg] \notag \\
&=\sum_{\bm\sigma\in\Omega} \prod_{i=1}^N\,
\exp\Bigg[
\beta g b \sigma_i
-\dfrac{\beta g}{2}N b^2
+\psi\eta_i^1 \sigma_i
+\tilde\psi \sigma_i
+A J_i \sigma_i
\Bigg] \notag \\
&= \exp \left( -\dfrac{\beta g}{2} N b^2\right)\prod_{i=1}^N \sum_{\sigma_i \in \{-1, +1\}} \exp \left[ \sigma_i \left( \beta g b + \psi + \tilde \psi + A J_i\right)\right] \notag \\
&=\exp \left( -\dfrac{\beta g}{2} N b^2\right)\prod_{i=1}^N 2\cosh \left( \beta g b + \psi + \tilde \psi + A J_i\right).
\end{aligned}
\end{equation}
Since the role of each $J_i$ is the same for all index $i$, we can consider just one $J$ and, in this way, we have 
\begin{align}
    Z_{g,N}^{(P)}(\beta \vert \bm \xi^1, \bm \lambda, \bm J; t=0)&=\exp \left( -\dfrac{\beta g}{2} N b^2\right) 2^N \cosh^N\left( \beta g b + \psi + \tilde \psi + A J\right). 
    \label{eq:Z0_app}
\end{align}
Inserting \eqref{eq:Z0_app} in the definition of the interpolating quenched statistical pressure \eqref{eq:stat-pres-interpolating} and applying the properties of the logarithm we get the desired one body terms. 

\section{Null-temperature limit computations}
\label{app:Tnullo}

Inspired by Amit's approach in \cite{Amit}, let us start by introducing the term $\beta y$ in the argument of the hyperbolic tangent in the self-consistency equations 
\eqref{eq:self_n_T_zero}, 
one obtains:
\begin{equation}
\label{eq:self_n_app}
\begin{aligned}
\bar{\tilde{m}}
&=
\mathbb{E}_{x,\xi^1}\Bigg\{
\big(\xi^1-b\big)
\tanh\Bigg[
\beta \left(\dfrac{P}{2}\bar{\tilde m}^{P-1}\big(\xi^1-b\big)
+ g (b-\bar{M})
+x\sqrt{\dfrac{\gamma P}{2}(1-b^2)^{P}\,\bar{q}^{P-1}} + y
\right)\Bigg]
\Bigg\},
\\
\bar{q}
&=
\mathbb{E}_{x,\xi^1}\Bigg\{
\tanh^2\Bigg[
\beta \left(\dfrac{P}{2}\bar{\tilde m}^{P-1}\big(\xi^1-b\big)
+ g (b-\bar{M})
+x\sqrt{\dfrac{\gamma P}{2}(1-b^2)^{P}\,\bar{q}^{P-1}}
+y\right)\Bigg]
\Bigg\},
\\
\bar{M}
&=
\mathbb{E}_{x,\xi^1}\Bigg\{
\tanh\Bigg[
\beta \left(\dfrac{P}{2}\bar{\tilde m}^{P-1}\big(\xi^1-b\big)
+ g (b-\bar{M})
+x\sqrt{\dfrac{\gamma P}{2}(1-b^2)^{P}\,\bar{q}^{P-1}}
\right)\Bigg]
\Bigg\}.
\end{aligned}
\end{equation}
where \(x\) is a standard Gaussian variable.  
We observe that  in the limit \(T \to 0\) (or equivalently \(\beta \to \infty\)), we get $\q \to 1$, allowing us to introduce the reparameterization 
\begin{align}
    \q=1-\dfrac{\delta \q}{\beta}, \qquad \beta \to +\infty.
\end{align}

Therefore, we get 
\begin{equation}
\begin{aligned}
\bar{\tilde{m}}
&=
\mathbb{E}_{x,\xi^1}\Bigg\{
\big(\xi^1-b\big)
\tanh\Bigg[
\beta \left(\dfrac{P}{2}\bar{\tilde m}^{P-1}\big(\xi^1-b\big)
+ g (b-\bar{M})
+x\sqrt{\dfrac{\gamma P}{2}(1-b^2)^{P}\,\bar{q}^{P-1}} + y
\right)\Bigg]
\Bigg\},
\\
1-\dfrac{\delta \q}{\beta}
&=
\mathbb{E}_{x,\xi^1}\Bigg\{
\tanh^2\Bigg[
\beta \left(\dfrac{P}{2}\bar{\tilde m}^{P-1}\big(\xi^1-b\big)
+ g (b-\bar{M})
+x\sqrt{\dfrac{\gamma P}{2}(1-b^2)^{P}\,\bar{q}^{P-1}}
+y\right)\Bigg]
\Bigg\},
\\
\bar{M}
&=
\mathbb{E}_{x,\xi^1}\Bigg\{
\tanh\Bigg[
\beta \left(\dfrac{P}{2}\bar{\tilde m}^{P-1}\big(\xi^1-b\big)
+ g (b-\bar{M})
+x\sqrt{\dfrac{\gamma P}{2}(1-b^2)^{P}\,\bar{q}^{P-1}}
\right)\Bigg]
\Bigg\}.
\end{aligned}
\end{equation}
Now we highlight that if we derive the equation for $\m$ with respect to $y$ we get 
\begin{align}
    \dfrac{\partial \m}{\partial y} = \beta \left[ 1-\left( 1-\dfrac{\delta \q}{\b }\right)\right] = \delta \q, 
\end{align}
namely
\begin{equation}
\label{eq:corresponding}
\begin{aligned}
\bar{\tilde{m}}
&=
\mathbb{E}_{x,\xi^1}\Bigg\{
\big(\xi^1-b\big)
\tanh\Bigg[
\beta \left(\dfrac{P}{2}\bar{\tilde m}^{P-1}\big(\xi^1-b\big)
+ g (b-\bar{M})
+x\sqrt{\dfrac{\gamma P}{2}(1-b^2)^{P}\,\bar{q}^{P-1}} + y
\right)\Bigg]
\Bigg\},
\\
\bar{M}
&=
\mathbb{E}_{x,\xi^1}\Bigg\{
\tanh\Bigg[
\beta \left(\dfrac{P}{2}\bar{\tilde m}^{P-1}\big(\xi^1-b\big)
+ g (b-\bar{M})
+x\sqrt{\dfrac{\gamma P}{2}(1-b^2)^{P}\,\bar{q}^{P-1}}
\right)\Bigg]
\Bigg\}, \\
\delta \q&= \dfrac{\partial \m }{\partial y}.
\end{aligned}
\end{equation}

For $\beta \to + \infty$ and for any $A,\ B$ constants we can say that,
\begin{align}
\int_{-\infty}^{\infty} dz \, 
\frac{\exp\!\left(-\tfrac{1}{2} z^{2}\right)}{\sqrt{2\pi}} 
\tanh\!\big[\beta(A z + B)\big]
= \int_{-\infty}^{\infty} dz \, 
\frac{\exp\!\left(-\tfrac{1}{2} z^{2}\right)}{\sqrt{2\pi}} 
\textnormal{sign} \!\big[\beta(A z + B)\big]=\operatorname{erf}\!\left(\dfrac{B}{\sqrt{2}A}\right). \label{eq:a}
\end{align}

Therefore, this means for $\beta \to +\infty$
\begin{align}
    \bar{M} =& \mathbb{E}_{\eta^1}\!\left[ 
    \textnormal{erf}\!\left( \dfrac{\dfrac{P}{{2}}\, \bar{\tilde m}^{P-1} (\xi^1-b) 
    +g(b-\bar{M} ) + y}{\sqrt{\gamma P(1-b^2)^P\,}}
   \right)\right]\\
   \m =& \mathbb{E}_{\eta_1}\!\left[
    \eta^1\, \textnormal{erf}\!\left( \dfrac{\dfrac{P}{{2}}\, \bar{\tilde m}^{P-1} (\xi^1-b) 
    +g(b-\bar{M} ) + y}{\sqrt{\gamma P(1-b^2)^P\,}}
   \right)\right]\\
   \q =&1
\end{align}
Putting $y=0$ and expanding the average with respect to $t$ completes the proof. 
\end{document}